\documentclass{ws-ijmpd}

\newcommand{\beqs}{\begin{equation*}}
\newcommand{\beq}{\begin{equation}}

\newcommand{\eeqs}{\end{equation*}}
\newcommand{\eeq}{\end{equation}}

\newcommand{\beqas}{\begin{eqnarray*}}
\newcommand{\beqa}{\begin{eqnarray}}

\newcommand{\eeqas}{\end{eqnarray*}}
\newcommand{\eeqa}{\end{eqnarray}}

\newcommand{\eq}[2]{\begin{equation} #1 \label{#2} \end{equation}}

\newcommand{\eps}{\varepsilon}
\newcommand{\al}{\alpha}
\newcommand{\be}{\beta}
\newcommand{\ga}{\gamma}
\newcommand{\de}{\delta}

\newcommand{\la}{\lambda}
\newcommand{\si}{\sigma}

\newcommand{\Ga}{\Gamma}

\newcommand{\blist}{\begin{itemize}}

\newcommand{\elist}{\end{itemize}}

\providecommand{\href}[2]{#2}

\DeclareFontFamily{OT1}{rsfs}{}
\DeclareFontShape{OT1}{rsfs}{m}{n}{ <-7> rsfs5 <7-10> rsfs7 <10->rsfs10}{} 
\DeclareMathAlphabet{\mycal}{OT1}{rsfs}{m}{n}

\DeclareMathOperator{\extdm}{d}
\newcommand{\extd}{\extdm \!}

\newcommand{\Rtr}{\hat{r}}
\newcommand{\Ftr}{\hat{t}}

\newcommand{\dual}[1]{\tilde{#1}}

\begin{document}

\markboth{D.~GRUMILLER AND R.~JACKIW}
{KALUZA-KLEIN REDUCTION OF CONFORMALLY FLAT SPACES}

\catchline{}{}{}{}{}

\title{KALUZA-KLEIN REDUCTION OF CONFORMALLY FLAT SPACES}

\author{D.~GRUMILLER}

\address{Institute for Theoretical Physics, University of Leipzig, \\
Augustusplatz 10-11, D-04109 Leipzig, Germany\\ and \\ 
Center for Theoretical Physics,
Massachusetts Institute of Technology,\\
77 Massachusetts Ave.,
Cambridge, MA  02139\\
grumil@lns.mit.edu}

\author{R.~JACKIW}

\address{Center for Theoretical Physics,
Massachusetts Institute of Technology,\\
77 Massachusetts Ave.,
Cambridge, MA  02139\\
jackiw@lns.mit.edu}

\maketitle

\begin{history}
LU-ITP 2006/14, MIT-CTP 3763, {\tt math-ph/0609025}

\received{11.~September 2006}
\accepted{21.~November 2006}
\comby{D.V.~Ahluwalia-Khalilova}
\end{history}

\begin{abstract}
Kaluza-Klein reduction of conformally flat spaces is considered for arbitrary dimensions. The corresponding equations are particularly elegant for the reduction from four to three dimensions. Assuming circular symmetry leads to explicit solutions which also arise from specific two-dimensional dilaton gravity actions.
\end{abstract}

\keywords{Conformally flat spaces; Kaluza-Klein reduction; 2D dilaton gravity.}

\section{Introduction}

A ``conformal tensor'' is constructed from the metric tensor $g_{MN}$ (or {\em Vielbein} $e_M^A$) and is invariant against Weyl rescaling $g_{MN}\to e^{2\si}g_{MN}$ (or $e_M^A\to e^\si e_M^A$). Moreover, it vanishes if and only if the space is conformally flat, $g_{MN}=e^{2\si}\eta_{MN}$ (or $e_M^A=e^\si\de^A_M$). In dimension four or greater the conformal tensor is the Weyl tensor. In three dimensions the Weyl tensor vanishes identically, while the Cotton tensor takes the role of probing conformal flatness and Weyl invariance.

Of interest is the behavior of the $n$-dimensional conformal tensor under a Kaluza-Klein dimensional reduction to $n-1$ dimensions. Specifically in $n$ dimensions we take the metric tensor in the form
\begin{equation}
  \label{eq:j1}
  g_{MN}=e^{2\si}\left(\begin{array}{cc}
  g_{\mu\nu}-a_\mu a_\nu & -a_\mu \\
  -a_\nu & -1  
   \end{array}\right)\,.
\end{equation}
Equivalently, for the {\em Vielbein} we take
\begin{equation}
  \label{eq:j2}
  e_M^A = e^\si \left(\begin{array}{cc}
  e_\mu^a & a_\mu \\
  0 & 1  
   \end{array}\right)\,.
\end{equation}
Here $g_{\mu\nu}$ is the $n-1$ dimensional metric tensor, $e_\mu^a$ is the corresponding {\em Vielbein}, $a_\mu$ is an $n-1$ dimensional vector. The scalar $\si$ plays no role in a conformal tensor and is therefore omitted in the following. It is assumed that the remaining quantities are independent of the $n^{\rm th}$ coordinate $x^n$. An $x^n$-independent redefinition of that coordinate acts as a gauge transformation on $a_\mu$. We use capital letters for $n$-dimensional indices, moved by $g_{MN}$, and Greek letters for $n-1$ dimensional indices, moved by $g_{\mu\nu}$. 
The $n-1$ dimensional line element 
\begin{equation}
  \label{eq:le1}
  \extd s^2_{(n-1)}=g_{\mu\nu}\extd x^\mu\extd x^\nu
\end{equation}
enters into the $n$-dimensional line element as
\begin{equation}
  \label{eq:le2}
  \extd s^2_{(n)}=g_{MN}\extd x^M\extd x^N=\extd s^2_{(n-1)}-(a_\mu\extd x^\mu+\extd x^n)^2\,.
\end{equation}
Tangent space indices are raised and lowered by the Minkowski metric with signature $(+,-,\dots,-)$. All $n-1$ dimensional geometric entities are denoted by lower case letters, whereas the $n$-dimensional Riemann tensor and related tensors are denoted by capital letters. 

Setting the dimensionally reduced $n$-dimensional conformal tensor to zero determines how our $n-1$ dimensional Kaluza-Klein theory fits into an $n$-dimensional, conformally flat space. By exhibiting specific forms for $g_{\mu\nu}$ and $a_\mu$, we provide a classification of conformally flat spaces with (at least) one (conformal) Killing vector.

The above program has been carried out already for the $n=3\to n=2$ transition, with the Cotton tensor undergoing the reduction \cite{Guralnik:2003we}. The present work deals with the general case: $n\to n-1$, $n\geq 4$, and also examines in greater detail the $n=4\to n=3$ reduction, which is especially intriguing since the 4-dimensional Weyl tensor produces a vanishing Weyl tensor in $n=3$. For comparison with previous results, we first record some of the formulas of the $n=3\to n=2$ analysis.

The 2-dimensional quantities, which descend from the 3-dimensional Cotton tensor with the Kaluza-Klein {\em Ansatz} \eqref{eq:j1} or \eqref{eq:j2}, are expressed in terms of the metric tensor $g_{\mu\nu}$, its Ricci curvature scalar $r$, and the field strength $f$ constructed from $a_\mu$ ($\eps^{01}=1$),
\begin{equation}
  \label{eq:j3}
  \partial_\mu a_\nu -\partial_\nu a_\mu := \sqrt{-g}\eps_{\mu\nu} f\,.
\end{equation}
The equations that ensure vanishing Cotton tensor evaluate the Ricci scalar in terms of $f^2$ and a constant $c$,
\begin{equation}
  \label{eq:j4}
  r=3f^2-c
\end{equation}
(some signs differ from previous papers owing to different conventions; here $R_{MN}:=\partial_K\Ga^K{}_{MN} - \dots$). Also, $f$ satisfies a ``kink'' equation ($d^2:=d_\mu d^\mu$)
\begin{equation}
  \label{eq:j5}
  d^2 f - cf+f^3=0
\end{equation}
and a traceless equation
\begin{equation}
  \label{eq:j6}
  d_\mu d_\nu f -\frac12 g_{\mu\nu} d^2 f = 0\,.
\end{equation}
Here $d$ is the lower-dimensional covariant derivative. Because the Cotton tensor arises by varying the 3-dimensional gravitational Chern-Simons term, the above equations also arise from an action, which is the dimensional reduction of the Chern-Simons term. This action takes the form of a 2-dimensional dilaton gravity
\begin{equation}
  \label{eq:j7}
  I_{\rm CS} = \frac{1}{8\pi^2}\int \extd^2x\sqrt{-g}\left(fr-f^3\right)\,,
\end{equation}
except that $f$ is not a fundamental dilaton field, but is the curl of $a_\mu$, \eqref{eq:j3}. Equation \eqref{eq:j4} comes from varying $I_{\rm CS}$ with respect to $a_\mu$, while \eqref{eq:j5} and \eqref{eq:j6} are obtained by varying $g_{\mu\nu}$. A reformulation of \eqref{eq:j7} as genuine Maxwell-dilaton gravity action is possible as well \cite{Grumiller:2003ad},
\begin{equation}
  \label{eq:d1}
  I_{\rm CS}^\prime = \frac{1}{8\pi^2}\int \extd^2x\sqrt{-g}\left(Xr - Yf + YX-X^3\right)\,,
\end{equation}
and facilitates the construction of all classical solutions. The dilaton $X$ is now a fundamental scalar field, which coincides with $f$ on-shell.

\section{Dimensional Reduction from $n$ to $n-1$ Dimensions}

Evaluating the $n$-dimensional Riemann tensor $R^{KLMN}$ on an $x^n$-independent metric tensor of the form \eqref{eq:j1} gives
\begin{subequations}
\label{eq:j21}
\begin{align}
R^{\mu\nu\la\tau}&=r^{\mu\nu\la\tau}+\frac14\left(f^{\mu\tau}f^{\la\nu}-f^{\mu\la}f^{\tau\nu}+2f^{\mu\nu}f^{\la\tau}\right) \label{eq:j21a} \\
R^{-\,\la\mu\nu}&=\frac12 d^\la f^{\mu\nu}-a_\tau R^{\tau\la\mu\nu} \label{eq:j21b} \\
R^{-\,\mu-\,\nu}&=-\frac14 f^{\mu\la}f_\la{}^\nu-a_\la(R^{-\mu\la\nu}+R^{-\nu\la\mu})-a_\la a_\tau R^{\la\mu\tau\nu} \label{eq:j21c}
\end{align} 
\end{subequations}
Here $R^{\mu\nu\la\tau}$ is the $n$-dimensional Riemann tensor with all indices evaluated in the $n-1$ dimensional range; in $R^{-\la\mu\nu}$ the first index refers to the $n^{\rm th}$ dimension, and this is similarly the case for $R^{-\mu-\nu}$; $r^{\mu\nu\la\tau}$ is the $n-1$ dimensional Riemann tensor and
\begin{equation}
  \label{eq:j22}
  f_{\mu\nu}:=\partial_\mu a_\nu-\partial_\nu a_\mu\,.
\end{equation}
The $n$-dimensional Ricci tensor has components
\begin{subequations}
\label{eq:j23}
\begin{align}
R^{\mu\nu}&=r^{\mu\nu}-\frac12 f^\mu{}_\la f^{\la\nu} \label{eq:j23a} \\
R^{-\,\mu}&= -\frac12 d_\nu f^{\nu\mu}-a_\nu R^{\mu\nu} \label{eq:j23b} \\
R^{-\,-}&=-\frac14 f^{\mu\nu}f_{\nu\mu}-2a_\mu R^{-\,\mu}-a_\mu a_\nu R^{\mu\nu} \label{eq:j23c}
\end{align}
\end{subequations}
Finally, the Ricci scalar reduces as
\begin{equation}
  \label{eq:j24}
  R=r-\frac14 f^{\mu\nu}f_{\nu\mu}\,.
\end{equation}
Of course \eqref{eq:j23} and \eqref{eq:j24} follow from \eqref{eq:j21} by taking the appropriate traces. These formulas are equivalent to the Gauss-Codazzi equations, which however are usually presented in a different manner.

Next we employ the formula that expresses the Weyl tensor $C^{KLMN}$ in terms of the Riemann tensor and its traces in $n$ dimensions 
\begin{subequations}
\label{eq:j25}
\begin{equation}
C^{KLMN} := R^{KLMN}-\frac{2}{n-2} \left(g^{K[M}S^{N]L}-g^{L[M}S^{N]K}\right) \label{eq:j25a}
\end{equation}
where $g^{K[M}S^{N]L}:=(g^{KM}S^{NL}-g^{KN}S^{ML})/2$ denotes antisymmetrization and 
\begin{equation}
S^{NL}:=R^{NL}-\frac{1}{2(n-1)}g^{NL}R \label{eq:j25b}
\end{equation}
\end{subequations}
is the Schouten tensor.
A similar formula holds in $n-1$ dimensions, with $n-1$ replacing $n$. We express the various curvatures in \eqref{eq:j25} in terms of the $n-1$ dimensional expressions in \eqref{eq:j21}-\eqref{eq:j24}, and then reexpress $r^{\mu\nu\la\tau}$ in terms of the $n-1$ dimensional Weyl tensor, using \eqref{eq:j25} with $n$ replaced by $n-1$. This gives our principal result of this Section: the relation between the $n$- and $n-1$ dimensional Weyl tensors, when the metric tensor is represented as in \eqref{eq:j1} 
\begin{subequations}
\label{eq:j26}
\begin{multline}
  \label{eq:j26a}
  C^{\mu\nu\la\tau} = c^{\mu\nu\la\tau} + \frac{2}{n-3}\left(g^{\mu[\la}c^{\tau]\nu}-g^{\nu[\la}c^{\tau]\mu}\right)\\
+\frac14\left(f^{\mu\tau}f^{\tau\nu}-f^{\mu\la}f^{\tau\nu}+2f^{\mu\nu}f^{\la\tau}\right)
+\frac{3}{2(n-3)}\left(g^{\mu[\la}t^{\tau]\nu}-g^{\nu[\la}t^{\tau]\mu}\right)
\end{multline}
where
\begin{align}
c^{\mu\nu}&:=\frac{1}{n-2}\left(r^{\mu\nu}-\frac{1}{n-1}g^{\mu\nu} r-\frac n4 \left(f^{\mu\la}f_\la{}^\nu-\frac{1}{n-1}g^{\mu\nu}f^{\la\tau}f_{\tau\la}\right)\right)\label{eq:j26b}\\
t^{\mu\nu} &:= f^{\mu\la}f_\la{}^\nu-\frac{1}{2(n-2)}g^{\mu\nu}f^{\la\tau}f_{\tau\la}
\label{eq:j26c}
\end{align}
Note that $c^{\mu\nu}=C^{\la\mu}{}_\la{}^\nu$ and $c^\mu{}_\mu=0$ (because only the traceless part of the Ricci tensor and a traceless combination of field strengths enters $c^{\mu\nu}$). The remaining independent component is
\begin{equation}
  \label{eq:j26d}
  C^{-\la\mu\nu}=\frac12 d^\la f^{\mu\nu} + \frac{1}{(n-2)} g^{\la[\mu}d_\tau f^{\nu]\tau}-a_\tau C^{\tau\la\mu\nu}\,.
\end{equation}
\end{subequations}
The further quantity $C^{-\mu-\nu}$ is determined by the previous, because $C^{KLMN}$ is traceless in all paired indices.

Our formulas \eqref{eq:j21}-\eqref{eq:j26} may also be presented with tangent space components, after contraction with {\em Vielbeine}. The only change is that the gauge-dependent, $a_\mu$-dependent contributions are absent.

The result \eqref{eq:j26} obviously makes sense only for $n\geq 4$; at $n=4$, $c^{\mu\nu\la\tau}$ --- the 3-dimensional Weyl tensor --- vanishes. Other simplifications occur as well, and this will be described below.

\section{Equations for Conformal Flatness}

\subsection{General $n\geq 4$}

Demanding that the $n$-dimensional Weyl tensor \eqref{eq:j26} vanish, requires that $C^{\mu\nu\la\tau}$ and $C^{-\la\mu\nu}$ vanish. Thus a $n$-dimensional conformally flat space obeys after a Kaluza-Klein reduction the following equations in $n-1$ dimensions
\begin{subequations}
\label{eq:j31}
\begin{align}
& c^{\mu\nu\la\tau}+\frac 14 \left(f^{\mu\tau}f^{\la\nu}-f^{\mu\la}f^{\tau\nu}+2f^{\mu\nu}f^{\la\tau}\right) +\frac{3}{2(n-3)}\left(g^{\mu[\la}t^{\tau]\nu}-g^{\nu[\la}t^{\tau]\mu}\right) = 0 \label{eq:j31a} \\
& c^{\mu\nu} = 0 \rightarrow r^{\mu\nu}-\frac{1}{n-1}g^{\mu\nu} r = \frac n4 \left(f^{\mu\la}f_\la{}^\nu-\frac{1}{n-1}g^{\mu\nu}f^{\la\tau}f_{\tau\la}\right) \label{eq:j31b} \\
& d^\la f^{\mu\nu} + \frac{2}{n-2} g^{\la[\mu}d_\tau f^{\nu]\tau} = 0 \label{eq:j31c}
\end{align}
\end{subequations}
The first two are traceless in all paired indices; the last one is consistent with the gauge theoretic Bianchi identity.

A general result consequent to \eqref{eq:j31b} and \eqref{eq:j31c} expresses the scalar curvature $r$ in terms of $f^{\mu\nu}f_{\nu\mu}$. This is established in the following manner. Take the covariant divergence of $c^{\mu\nu}$ and use the Bianchi identity $d^\mu r_{\mu\nu}=\frac12 \partial_\nu r$. This gives
\begin{subequations}
\label{eq:j32}
\begin{equation}
  \label{eq:j32a}
  \partial_\nu\left(2(n-3)r+nf^{\la\tau}f_{\tau\la}\right)=n(n-1)\left(d_\mu f^{\mu\la}f_{\la\nu}+f^{\mu\la}d_\mu f_{\la\nu}\right)\,.
\end{equation}
The last term is evaluated from \eqref{eq:j31c} as
\begin{equation}
  \label{eq:j32b}
  f^{\mu\la}d_\mu f_{\la\nu} = \frac{1}{n-2} d_\mu f^{\mu\la}f_{\la\nu}
\end{equation}
and combines with the first term on the right of \eqref{eq:j32a} to give $\frac{n(n-1)^2}{n-2}d_\mu f^{\mu\la}f_{\la\nu}$. This is now shown to be a total derivative. To accomplish the result, begin with the equality that follows from the Bianchi identity for $f_{\la\nu}$
\begin{equation}
  \label{eq:j32c}
  f^{\mu\la}d_\mu f_{\la\nu}=-f^{\mu\la}\left(d_\la f_{\nu\mu}+d_\nu f_{\mu\la}\right) = - f^{\mu\la}d_\mu f_{\la\nu}+\frac 12 \partial_\nu\left(f^{\mu\la}f_{\la\mu}\right)\,.
\end{equation}
Equivalently
\begin{equation}
  \label{eq:j32d}
  f^{\mu\la}d_\mu f_{\la\nu}=\frac14\partial_\nu\left(f^{\mu\la}f_{\la\mu}\right)
\end{equation}
But from \eqref{eq:j32b}, this implies
\begin{equation}
  \label{eq:j32e}
  d_\mu f^{\mu\la}f_{\la\nu} = \frac{n-2}{4}\partial_\nu\left(f^{\mu\la}f_{\la\mu}\right)\,.
\end{equation}
\end{subequations}
Therefore, the right side of \eqref{eq:j32a} is indeed a total derivative and its integration gives finally
\begin{equation}
  \label{eq:j33}
  r=\frac{n(n+1)}{8}f^{\mu\nu}f_{\nu\mu}-c\,,
\end{equation}
where $c$ is a constant. This is a generalization to arbitrary dimension of our $n=3$ result \eqref{eq:j4}. 

A further general, but trivial result may be derived. Equations \eqref{eq:j31} are solved by vanishing $f_{\mu\nu}$, pure gauge $a_\mu$, because then \eqref{eq:j31c} is true and \eqref{eq:j31b} shows that the space is maximally symmetric. Therefore the Weyl tensor vanishes, thereby satisfying \eqref{eq:j31a}. Finally, equation \eqref{eq:j33} consistently identifies the scalar curvature with a constant.

If viewed as a field equation, \eqref{eq:j31b} has a status comparable to the Einstein equation as it connects geometry (left hand side) with matter (right hand side). Incidentally, if $n=5$ the right hand side is proportional to the energy momentum tensor of a 4-dimensional Maxwell field, concurrent with the conformal properties of the latter in four dimensions. 

\subsection{$n=4$}

Our equations simplify dramatically when $n=4$ and the reduced system is 3-dimensional. First of all, the $n=3$ Weyl tensor vanishes identically. Moreover, the field strength $f_{\mu\nu}$ may be presented in terms of its dual $f^\mu$
\begin{equation}
  \label{eq:j34}
  f_{\mu\nu}=\sqrt{g}\eps_{\mu\nu\la} f^\la\,,\qquad d_\mu f^\mu=0\,.
\end{equation}
Substituting \eqref{eq:j34} into \eqref{eq:j26a} shows that the last two $f$-dependent quantities vanish identically, leaving
\begin{subequations}
\label{eq:j35}
\begin{align}
  \label{eq:j35a}
C^{\mu\nu\la\tau} & =2\left(g^{\mu[\la}c^{\tau]\nu}-g^{\nu[\la}c^{\tau]\mu}\right) \\
\label{eq:j35b}
c^{\mu\nu} & =\frac 12 \left(r^{\mu\nu}-\frac13 g^{\mu\nu}r-f^\mu f^\nu + \frac 13 g^{\mu\nu} f^2\right)
\end{align}
($f^2:=f^\mu f_\mu$), while \eqref{eq:j26d} becomes
\begin{equation}
  \label{eq:j35c}
  C^{-\la\mu\nu} = \frac{\eps^{\mu\nu\tau}}{2\sqrt{g}}\left(d^\la f_\tau + d_\tau f^\la\right)-a_\tau C^{\tau\la\mu\nu}
\end{equation}
\end{subequations}
Therefore the vanishing of the 4-dimensional Weyl tensor, when reduced to three dimensions, requires according to \eqref{eq:j35a} and \eqref{eq:j35b} that $c^{\mu\nu}$ vanish
\begin{subequations}
\label{eq:j36}
\begin{equation}
  \label{eq:j36a}
  r^{\mu\nu}-\frac{1}{3}g^{\mu\nu} r = f^\mu f^\nu-\frac13 g^{\mu\nu}f^2 
\end{equation}
while according to \eqref{eq:j35c} $f^\mu$, when it is non-vanishing, is a Killing vector of the 3-dimensional geometry
\begin{equation}
  \label{eq:j36b}
  d_\mu f_\nu + d_\nu f_\mu = f^\la\partial_\la g_{\mu\nu} + g_{\la\nu} \partial_\mu f^\la  + g_{\mu\la} \partial_\nu f^\la =0\,.
\end{equation}
\end{subequations}
The $n=4$ restriction of \eqref{eq:j33} is a straightforward consequence of equations \eqref{eq:j36},
\begin{equation}
  \label{eq:j36c}
  r=-5f^2-c\,.
\end{equation}
Equation \eqref{eq:j36a} may be presented in ``Einstein form''
\begin{equation}
  \label{eq:j37}
  r_{\mu\nu}-\frac12 g_{\mu\nu} r = f_\mu f_\nu+\frac16 g_{\mu\nu}\left(c+3 f^2\right)
\end{equation}
For time-like $f^\mu$ this has a hydrodynamical interpretation with pressure $P=-\frac c6-\frac12 f^2$ and energy density $E=\frac32 f^2+\frac c6$.

When $f^\mu$ vanishes \eqref{eq:j36a} shows that our 3-dimensional space is maximally symmetric. This gives the line element
\begin{subequations}
\begin{equation}
  \label{eq:le3}
  \extd s^2_{(4)}=\extd s^2_{(3)}-\extd z^2
\end{equation}
where $z=x^4$ and $\extd s^2_{(3)}$ describes 
Minkowski or $(A)dS_3$ space,
\begin{equation}
  \label{eq:le4}
  \extd s^2_{(3)}=(1\mp\la^2\rho^2)\extd t^2-(1\mp\la^2 \rho^2)^{-1}\extd \rho^2-\rho^2\extd\theta^2\,.
\end{equation}
\end{subequations}
The upper (lower) sign is valid for $(A)dS_3$. Minkowski space is obtained if the parameter $\la$ vanishes. The Ricci scalar is given by $r=\mp 6\la^2$.
Of course
the 4-dimensional line element \eqref{eq:le3} is conformally flat by construction. 

For non-vanishing $f^\mu$, the right side of \eqref{eq:j36a} must not vanish, for otherwise the metric would be singular. Therefore, with non-vanishing $f^\mu$ the space is not maximally symmetric. In that case a second Killing vector exists if $d_\mu f_\nu$ does not vanish identically. This statement can be derived as follows. The relation $d_\mu d_\nu f_\la = R_{\la\nu\mu\tau}f^\tau$  --- upon expressing the Riemann tensor in terms of the Ricci tensor and the latter in terms of the Killing vector using \eqref{eq:j36a} and \eqref{eq:j36c} --- establishes
\begin{subequations}
\begin{equation}
  \label{eq:2ndkill1}
  d_\mu d_\nu f_\la = \frac16\left(g_{\mu\nu}f_\la-g_{\mu\la}f_\nu\right)\left(3f^2+c\right)\,.
\end{equation}
Defining
\begin{equation}
  \label{eq:2ndkill2}
  F^\mu := \frac{\eps^{\mu\nu\la}}{\sqrt{g}}d_\nu f_\la
\end{equation}
leads by virtue of \eqref{eq:2ndkill1} to the Killing equation for $F^\mu$,
\begin{equation}
  \label{eq:2ndkill3}
  d_\mu F_\nu + d_\nu F_\mu = 0\,.
\end{equation}
\end{subequations}
We say that $F^\mu$ is a Killing vector dual to the Killing vector $f^\mu$, which enters our system of equations. 
It then remains a further problem whether this dual Killing vector
determines an (additional) geometry from \eqref{eq:j36}. 
(The dual Killing vector $F^\mu$ must not be confused with the gauge
theoretic dual of $f^\mu$, viz.~$f_{\mu\nu}=\sqrt{g}\eps_{\mu\nu\la}f^\la$.)

If $d_\mu f_\nu=0$ (but $f^\mu\neq 0$) then there is no dual Killing
vector of the form \eqref{eq:2ndkill2}. 
From \eqref{eq:2ndkill1} one may deduce $3f^2+c=0$ which together with \eqref{eq:j36c} leads to constant curvature $r=-2f^2$. The Killing condition \eqref{eq:j36b} is fulfilled trivially and \eqref{eq:j36a} or \eqref{eq:j37} simplify to
\begin{equation}
  \label{eq:covconst1}
  r_{\mu\nu}-\frac12 g_{\mu\nu}r = f_\mu f_\nu\,.
\end{equation}
(In a hydrodynamical interpretation this is the Einstein equation with a
pressureless perfect fluid source of constant energy density if $f^\mu$
is time-like.) We shall not analyze this case further here but encounter
it again in Section \ref{se:CDV} below.

While our equations \eqref{eq:j36} posses an undeniable elegance, we have not succeeded in analyzing them further in full generality. Due to the presence of a Killing vector $f^\mu$ one can perform a second Kaluza-Klein split \eqref{eq:j1} for the 3-dimensional metric and study the ensuing equations. In the next Section we present all solutions with vanishing vector potential (with respect to the second Kaluza-Klein split), i.e., the 3-dimensional Killing vector is required to be hypersurface orthogonal and the metric tensor is block diagonal.
With the notable exception of these solutions, we have not found an action that would generate \eqref{eq:j36}, since the 4-dimensional Weyl tensor is not known to be the variation of any action.

\section{Circularly Symmetric Solutions}

When $n>4$, the equations \eqref{eq:j31} determining the geometry and the field strength are daunting, and we have not attempted to solve them. For $n=4$ we have not found the general solution to \eqref{eq:j36}, but we have constructed special solutions, based on an {\em Ansatz} for the 3-dimensional line element similar to \eqref{eq:j1},
\begin{equation}
  \label{eq:dd}
  \extd s^2_{(3)} = g_{\al\be} \extd x^\al \extd x^\be - \phi^2 \extd\theta^2\,,
\end{equation}
but with vanishing vector field for simplicity. The 2-dimensional metric $g_{\al\be}$ and the scalar field $\phi$ are required to depend solely on the 2-dimensional coordinates $x^\al$, so we have circular symmetry due to the Killing vector $(\partial_\theta)^\mu$. 
A motivation for this {\em Ansatz} is that it encompasses the circularly symmetric case $\phi=\rho$, $x^\al=(t,\rho)$.
The 3-dimensional Riemann and Ricci tensors, as well as the Ricci scalar can be related to intrinsically 2-dimensional (2D) quantities, which are denoted in capital letters, with superscript ${(2)}$.\\
Riemann tensor:
\begin{subequations}
\label{eq:DD1}
\begin{align}
r^{\al\be\ga\de}&= {^{(2)\!}R^{\al\be\ga\de}}\,,\\
r^{\theta\al\be\ga}&=0\,,\\
r^{\theta\al\theta\be}&=\frac{1}{\phi^2}\left(\nabla^\be\nabla^\al\ln{\phi}+(\nabla^\al\ln{\phi})(\nabla^\be\ln{\phi})\right)\,,
\end{align}
Ricci tensor:
\begin{align}
r^{\alpha\beta}&= {^{(2)\!}R^{\al\be}}-\nabla^{\al}\nabla^{\be}\ln{\phi}-(\nabla^\al\ln{\phi})(\nabla^\be\ln{\phi})\,,\\
r^{\al\theta}&=0\,,\\
r^{\theta\theta}&=\frac{1}{\phi^2}\left(\square\ln{\phi}+(\nabla\ln{\phi})^2\right)\,,
\end{align}
Ricci scalar:
\eq{
r={^{(2)\!}R}-2\left(\square\ln{\phi}+(\nabla\ln{\phi})^2\right)
}{mit:new18}
\end{subequations}
The indices $\al,\be,\ga,\de$ from the beginning of the Greek alphabet range over $(0,1)$ and $\theta$ denotes the Killing coordinate. The 2D covariant derivative is denoted by $\nabla_\al$ and $\square:=\nabla^\al\nabla_\al$.

We seek now solutions to \eqref{eq:j36} with a line element of the form \eqref{eq:dd}. First we check the Killing condition \eqref{eq:j36b} which ramifies into
\begin{subequations}
\label{eq:DD2}
\begin{align}
\nabla_{(\al} f_{\be)} &= 0 & \rightarrow & &f^\la\partial_\la g_{\al\be} + g_{\la\be}\partial_\al f^\la + g_{\la\al}\partial_\be f^\la &=0 \,,\label{kill1}\\
\nabla_{(\al} f_{\theta)} &= 0 & \rightarrow & &-\phi^2\partial_\al f^\theta + g_{\al\be}\partial_\theta f^\be &=0 \,,\label{kill2}\\
\nabla_\theta f_\theta &= 0 & \rightarrow & &\partial_\theta f^\theta + f^\al\partial_\al\ln{\phi} &=0 \,.\label{kill3}
\end{align}
\end{subequations}
Next we take note of the condition \eqref{eq:j36a} that $c^{\mu\nu}$ vanish. The five independent components split into the $\al\theta$-part
\begin{subequations}
\label{eq:const}
\begin{equation}
  \label{eq:const1}
  f^\al f^\theta = 0
\end{equation}
and the $\al\be$-part
\begin{equation}
  \label{eq:const2}
  r^{\al\be}-\frac{1}{3}g^{\al\be} r = f^\al f^\be-\frac13 g^{\al\be}\left(f^\ga f_\ga+f^\theta f_\theta\right)\,.
\end{equation}
\end{subequations}
The $\theta\theta$-part is redundant because $c^{\mu\nu}$ is traceless. 
Equation \eqref{eq:const1} requires either $f^\theta=0$, so the Killing vector would be intrinsically 2D, or $f^\al=0$, so the Killing vector would have no 2D component at all. A mixing, $f^\theta\neq 0\neq f^\al$, is not possible. So if $f^\al\neq 0$ --- and thus $f^\theta=0$ --- we deduce from \eqref{kill2} that $f^\al$ must be independent of $\theta$ and from \eqref{kill3} that it must be orthogonal to $\partial_\al \phi$. On the other hand, if $f^\theta\neq 0$  --- and thus $f^\al=0$ --- we deduce from \eqref{kill2} and \eqref{kill3} that $f^\theta$ must be constant. Thus, the 3-dimensional Killing vector always must be $\theta$-independent.

\subsection{Solutions of constant $\phi$}\label{se:CDV}

Before studying the general problem we shall focus on the much simpler case of constant $\phi$, scaled to unity, whence $r={^{(2)\!}R}$ and $r_{\al\be}={^{(2)\!}R_{\al\be}}=\frac12 g_{\al\be} {^{(2)\!}R}$. 
For $f^\theta=0$ and $f^\al\neq 0$ there is no solution to \eqref{eq:const2}. For $f^\al=0$ and $f^\theta=\la=\rm constant$ all conditions \eqref{eq:DD2} and \eqref{eq:const} are satisfied provided the Ricci scalar obeys
\begin{equation}
  \label{eq:const4}
  r={^{(2)\!}R}=-2f^\theta f_\theta = 2\la^2 \geq 0\,.
\end{equation}
In our notation this implies $AdS_2$ (or 2D Minkowski space if the inequality is saturated). It should be noted that the constant $c$ in \eqref{eq:j36c} may not be chosen freely but rather is determined as $c=3\la^2$. 

Thus all solutions with $\phi$ in \eqref{eq:dd} constant are either 2D Minkowski space or
$AdS_2$ and therefore admit three Killing vectors in 2D which may be lifted to 3-dimensional ones, thereby supplementing the $f^\mu$ with which we have started. The 3-dimensional line element
\begin{subequations}
\label{eq:solcons}
\begin{equation}
  \label{eq:le6}
   \extd s^2_{(3)}=(1+\la^2\rho^2)\extd t^2-(1+\la^2\rho^2)^{-1}\extd \rho^2-\extd\theta^2
\end{equation}
enters the 4-dimensional one (using the gauge $a_t=-\la\rho$ and $a_\rho=a_\theta=0$)
\begin{equation}
  \label{eq:le7}
  \extd s^2_{(4)}=\extd s^2_{(3)}-(\la\rho\extd t - \extd z)^2
\end{equation}
which again
\end{subequations}
is conformally flat by construction. 
All four Killing vectors can be lifted to four dimensions (only in one case this is not entirely trivial as the corresponding Killing vector acquires a $z$-component).

Note that the dual
Killing vector from \eqref{eq:2ndkill2} vanishes identically. This
brings us back to the case $d_\mu f_\nu=0$ mentioned around \eqref{eq:covconst1}. 
What we have shown above is that for the {\em
Ansatz} \eqref{eq:dd} and constant $\phi$ the special case $d_\mu
f_\nu=0$ emerges. To show that the converse holds (for space-like
$f^\mu$) without loss of generality  we can make a Kaluza-Klein {\em Ansatz}  
\begin{subequations}
\begin{equation}
  \label{eq:explan1}
  \extd s^2_{(3)} = g_{\mu\nu}\extd x^\mu\extd x^\nu = e^{2\hat{\si}} \left(\hat{g}_{\al\be}\extd x^\al\extd x^\be-(\hat{a}_\al\extd x^\al + \extd\theta)^2 \right)\,, 
\end{equation}
where the coordinate $\theta$ corresponds to the Killing direction implied by $f^\mu$ and $\al,\be$ range from 0 to 1. Therefore, $f^\al=0$ and $f^\theta=\rm constant$. All quantities depend on $x^\al$ only. So far we have just exploited the fact that $f^\mu$ is a (space-like) Killing vector and therefore the metric may be brought into the adapted form \eqref{eq:explan1}; now we employ the property that $f^\mu$ is covariantly constant.  
As a consequence $\partial_\al g_{\theta\theta}=0$, so the scalar field
$\hat{\si}$ must be constant and $e^{2\hat{\si}}$ can be scaled to 1. Next, we evaluate \eqref{eq:covconst1}
with upper indices for the $\al\be$-part and obtain by virtue of
\eqref{eq:j23a} and \eqref{eq:j24} (now evaluated for the reduction from
$3\to 2$) 
\begin{equation}
\label{eq:insert0}
{^{(2)\!}R}^{\al\be}-\frac12 \hat{f}^\al{}_\ga\hat{f}^{\ga\be} - \frac12 \hat{g}^{\al\be}\left( {^{(2)\!}R} -\frac14 \hat{f}^{\ga\de}\hat{f}_{\de\ga}\right)=f^\al f^\be=0\,,
\end{equation}
where $\hat{f}_{\al\be}:=\partial_\al\hat{a}_\be-\partial_\be\hat{a}_\al$.
With the identity ${^{(2)\!}R}^{\al\be}=\frac12 \hat{g}^{\al\be} \,{^{(2)\!}R}$ and the definition $\sqrt{-g}\eps_{\al\be}\hat{f}:=\hat{f}_{\al\be}$ equation \eqref{eq:insert0} simplifies to
\begin{equation}
\label{eq:insert1}
\hat{g}^{\al\be}\hat{f}^2 = 0\,.
\end{equation}
\end{subequations}
 The only solution to \eqref{eq:insert1} with non-degenerate 2D metric $\hat{g}^{\al\be}$ is given by $\hat{f}=0$, so $\hat{a}_\al$ must be pure gauge.
Thus, the 2-dimensional vector potential $\hat{a}_\al$ may be chosen to
vanish, leading to \eqref{eq:dd} with $\phi=1$. 
We have noted above that all such solutions exhibit three Killing vectors in addition to $f^\mu$.

\subsection{Solutions of non-constant $\phi$}

According to the previous analysis, there are two cases to consider for the Killing vector $f^\mu=(f^\al,f^\theta)$ in our system of equations \eqref{eq:DD2}-\eqref{eq:const}: intrinsically 2D Killing vector ($f^\al\neq 0$, $f^\theta=0$) or Killing vector with no 2D component ($f^\al=0$, $f^\theta=\la=\rm constant$). We discuss each case in turn.

\subsubsection{Intrinsically 2D Killing vector ($f^\al\neq 0$, $f^\theta=0$)}

We may always bring the 2D portion of the line element \eqref{eq:dd} into Eddington-Finkelstein form
\eq{
g_{\al\be}\extd x^\al\extd x^\be = e^{Q(X)}\left(2\extd u\extd X+K(X)\extd u^2\right)\,.
}{eq:gj1}
The metric functions depend on only one variable because, by hypothesis, there exists the Killing vector $f^\al$, which according to \eqref{eq:gj1} is proportional to $(\partial_u)^\al$. In components $f^\al=(f^u=1,f^X=0)$, where no generality is lost by rescaling the constant $f^u$ to $1$.  The Killing conditions \eqref{eq:DD2} require that $\phi$ is a function of $X$ only. The presence in \eqref{eq:gj1} of the prefactor $e^{Q(X)}$ still allows arbitrariness in the choice of the $X$ coordinate, which we shall fix by setting $\phi^2=X$. Equation \eqref{eq:const1} is obviously satisfied, and the remaining three conditions encoded in \eqref{eq:const2} simplify to the following two:
\begin{subequations}
\label{eq:minimaster}
\eq{
e^{2Q}=\frac{1}{4X^2}+\frac{Q^\prime}{2X}
}{eq:gj8}
\eq{
K'' + K^\prime\left(Q^\prime-\frac{1}{2X}\right) + K\left(Q''-\frac{1}{2X}Q^\prime\right) =  0 \,.
}{eq:gj9}
\end{subequations}
To solve \eqref{eq:gj8}, differentiate that equation and eliminate the exponential, leaving a differential equation for $Q^\prime$ which is easily solved as
\begin{subequations}
\eq{
Q^\prime=-\frac{1}{2X}-\frac{1}{2(X-a)}\,,\qquad a\in\mathbb{R}\,.
}{eq:gj10}
We assume $0\leq X\leq a$ and rescale $X$ by $x=X/a$. Thus $e^{-Q}$ is proportional to $\sqrt{x}\sqrt{1-x}$. The proportionality constant is irrelevant and may be absorbed into a global redefinition of units of length. Therefore we set it to 1.
\eq{
e^{-Q} = \sqrt{x}\sqrt{1-x}\,.
}{eq:gj11}
The solution to \eqref{eq:gj9} then becomes
\eq{
K=e^{-Q}\left(A+B\sqrt{1-x}\right)\,,\qquad A,B\in\mathbb{R}\,,
}{eq:gj12}
\end{subequations}
where $A,B$ are integration constants. With
\eq{
\Phi=2\arcsin{\sqrt{x}}\,,\qquad (0\leq \Phi\leq\pi)
}{eq:nolabel}
the 3-dimensional line element reads
\begin{subequations}
\label{sol1}
\eq{
\extd s^2_{(3)}=2\extd u\extd\Phi +\extd u^2\left(A+B\cos{(\Phi/2)}\right)-a\sin^2{(\Phi/2)}\extd\theta^2\,.
}{eq:gj13}
The coordinate redefinition
\eq{
\tanh{(\dual{\Phi}/2)}=\sin{(\Phi/2)}
}{eq:gj13ab}
allows an alternative presentation of the line element \eqref{eq:gj13} as
\eq{
\extd s^2_{(3)}=\frac{1}{\cosh{(\dual{\Phi}/2)}}\left(2\extd u\extd\dual{\Phi}+\extd u^2\left(B+A\cosh{(\dual{\Phi}/2)}\right)\right)-a\tanh^2{(\dual{\Phi}/2)}\extd\theta^2\,.
}{eq:gj13aa}
Note that in a sense $A$ and $B$ have interchanged their roles. This
will be elaborated in the next Section.
The corresponding Ricci scalar,
\begin{equation}
  \label{eq:r1}
  r=-\frac{5B}{4}\cos{(\Phi/2)}-\frac{A}{2} = -\frac{5B}{4}\frac{1}{\cosh{(\dual{\Phi}/2)}}-\frac{A}{2}
\end{equation}
has no singularities. 
Upon redefining $t=u+h(\Phi)$, $h^\prime(\Phi)=1/(A+B\cos{(\Phi/2)})$ and $\rho=\sqrt{a}\sin{(\Phi/2)}$ the line element
\begin{equation}
  \label{eq:le9}
  \extd s^2_{(3)}=(A+B\sqrt{1-\rho^2/a})\extd t^2-\frac{4/a}{(1-\rho^2/a)} (A+B\sqrt{1-\rho^2/a})^{-1} \extd\rho^2-\rho^2\extd\theta^2
\end{equation}
is in a conventional static and circularly symmetric form and the relevant Killing vector is time-like ($f^t=1$, $f^\rho=f^\theta=0$). The 4-dimensional line element (in the gauge $a_t=a_\rho=0$ and $a_\theta=-2\sqrt{a-\rho^2}$)
\begin{equation}
  \label{eq:le10}
  \extd s^2_{(4)}=\extd s^2_{(3)}-\left(2\sqrt{a-\rho^2}\extd\theta-\extd z\right)^2
\end{equation}
\end{subequations}
is again conformally flat by construction. 

Note that the Killing vector $F^\mu$ dual to our $f^\mu:(f^u=1,f^\Phi=0,f^\theta=0)$ is given by
\begin{equation}
  \label{eq:newlabel1}
  F^\mu=\frac{\eps^{\mu\nu\la}}{\sqrt{g}}\partial_\nu (g_{\la \tau} f^\tau)= 
\frac{\eps^{\mu\Phi u}}{\sqrt{g}}\partial_\Phi g_{u u}\,.
\end{equation}
Evidently only the $\theta$ component survives in $F^\mu$ and it is a constant. This leads to the other solution to our problem, with $f^\al=0$, $f^\theta={\rm constant}\neq 0$.

\subsubsection{Intrinsically 2D dual Killing vector ($f^\al=0$, $f^\theta\neq 0$)}

Upon choosing the constant $f^\theta$ to be $1$, we find that the dual Killing vector
\begin{equation}
  \label{eq:vkv1}
  F^\mu=\frac{\eps^{\mu\nu\la}}{\sqrt{g}}\partial_\nu(g_{\la\tau}f^\tau)=\frac{\eps^{\mu \al \theta}}{\sqrt{g}}\partial_\al \phi^2
\end{equation}
possesses only $\al$ components, i.e., is an intrinsically 2D Killing vector. Without loss of generality we apply again the Eddington-Finkelstein form \eqref{eq:gj1} for the 2D line element. Further, we now fix $X=\phi$.

Again the conditions \eqref{eq:const2} lead to two independent equations that read
\begin{subequations}
\label{eq:newminimaster}
\begin{equation}
  \label{eq:vkv2}
  Q^\prime = 0
\end{equation}
\begin{equation}
  \label{eq:vkv3}
  K''-K^\prime\left(Q^\prime+\frac{1}{X}\right)=2e^Q X^2
\end{equation}
\end{subequations}
with solutions
\begin{subequations}
\label{eq:newminimaster1}
\begin{equation}
  \label{eq:vkv21}
  e^Q = 1\,,
\end{equation}
\begin{equation}
  \label{eq:vkv31}
  K=\frac14 X^4 + A X^2 + B\,,\qquad (A,B\in\mathbb{R})\,,
\end{equation}
\end{subequations}
where $A$ and $B$ are integration constants. The integration constant inherent to $Q$ again has been fixed conveniently and without loss of generality.
Therefore the 3-dimensional line element becomes
\begin{subequations}
\label{sol2}
\begin{equation}
  \label{eq:vkv17}
  \extd s^2_{(3)}=2\extd u\extd X + \left(\frac{1}{4}X^4+A X^2+B\right)\extd u^2-X^2\extd\theta^2\,.
\end{equation}
The corresponding Ricci scalar,
\begin{equation}
  \label{eq:r2}
  r=5 X^2+6A\,,
\end{equation}
is singular for $|X|\to\infty$. With the coordinate transformation $t=u+h(X)$, $h^\prime=1/K$ and $\rho=X$ the 3-dimensional line element
\begin{equation}
  \label{eq:le11}
  \extd s^2_{(3)}=(\frac14 \rho^4+A\rho^2+B)\extd t^2-(\frac14\rho^4+A\rho^2+B)^{-1}\extd\rho^2-\rho^2\extd\theta^2
\end{equation}
is again in a conventional static and circularly symmetric form and the relevant Killing vector is space-like ($f^t=f^\rho=0$, $f^\theta=1$). The 4-dimensional line element (in the gauge $a_\theta=a_\rho=0$ and $a_t=-\rho^2/2$)
\begin{equation}
  \label{eq:le12}
  \extd s^2_{(4)}=\extd s^2_{(3)}-\left(\frac12 \rho^2\extd t-\extd z\right)^2
\end{equation}
\end{subequations}
is of course again conformally flat by construction.

Finally we observe that the Killing vector dual to our $f^\mu: (f^\al=0,f^\theta=1)$ possesses only a $u$ component and also is constant. In this way we are brought back to the previous case: intrinsically 2D Killing vector.

\section{2D dilaton gravity} 

One may wonder whether the 2D part of the solutions \eqref{eq:gj13} and
\eqref{eq:vkv17} can be derived from some action principle. The purpose
of this Section is to show that the answer is affirmative. Let us collect the evidence obtained so far: the field content of our
3-dimensional theory comprises the 2D metric $g_{\al\be}$ and a scalar field $\phi$; one obtains maximally symmetric 2D spaces if the scalar field is constant; there is always (at least) one 2D Killing vector. Therefore, we may conjecture that there should be an effective description in terms of 2D dilaton gravity because it shares all these features \cite{Grumiller:2002nm}. Thus, we would like to investigate now whether the conditions \eqref{eq:j36} with the {\em Ansatz} \eqref{eq:dd} could follow as equations of motion from an action
\eq{
I=\frac{1}{8\pi^2}\int\extd^2x\sqrt{-g}\left(XR+U(X)(\nabla X)^2-V(X)\right)\,,
}{eq:gj5}
where the dilaton field $X$ is some function of $\phi$ and $U,V$ are arbitrary functions defining the model. 
Since there is no obvious way to obtain an action in four dimensions leading to $C^{KLMN}=0$ as equations of motion, such a construction is of particular interest. We have dropped the superscript $(2)$ of the 2D Ricci scalar $R$ in \eqref{eq:gj5} as from now on we work almost exclusively in 2D and no confusion with 4-dimensional quantities should arise. 

We recall now a basic result of 2D dilaton gravity \cite{Grumiller:2002nm}:
The generic solution to the equations of motion following from
\eqref{eq:gj5} is parameterized by one constant of motion, $M$, and
using the dilaton field $X$ as one of the coordinates leads to the line element \eqref{eq:gj1} with
\begin{subequations}
\label{eq:eqs}
\begin{align}
\label{eq:gj7}
Q(X)&=-\int^X U(y)\extd y\,,  \\
K(X)&=\int^X e^{Q(y)}V(y)\extd y \,. 
\label{eq:gj6}
\end{align}
\end{subequations}
The ambiguities in defining $Q$ and $K$ due to the integration constants
from the lower limits correspond to a simple coordinate redefinition of
$u$ and an arbitrary aforementioned constant $M$ in the solution (a ``modulus''),
respectively. 
[In addition there may be isolated solutions (in the sense of not possessing a modulus), so-called constant dilaton vacua, which have a constant dilaton $X$ solving the equation $V(X)=0$ and leading to maximally symmetric spaces with curvature $R=V'(X)$.]

\subsection{Intrinsically 2D Killing vector}

The 2D line element extracted from \eqref{eq:gj13}, with $\Phi$ replaced by $X$,
\begin{subequations}
\label{eq:2D}
\eq{
\extd s^2=2\extd u\extd X +\extd u^2\left(A+B\cos{(X/2)}\right)
}{eq:gj14}
follows from the 2D dilaton gravity action 
\eq{
I_1=\frac{1}{8\pi^2}\int\extd^2x\sqrt{-g}\left(X R + \frac B2 \sin{(X/2)}\right)\,,
}{eq:gj15}
where $B$ is a parameter of the action and $A$ emerges as a constant of motion. So comparison with \eqref{eq:gj5} establishes the potentials
\eq{
U(X)=0\,,\qquad V(X) = -\frac B2 \sin{(X/2)}\,.
}{eq:gj16}
\end{subequations}

The alternative presentation \eqref{eq:gj13aa}, with $\dual{\Phi}$ replaced by $\dual{X}$, gives the line element
\begin{subequations}
\eq{
\extd s^2=\frac{1}{\cosh{(\dual{X}/2)}}\left(2\extd u\extd\dual{X}+\extd u^2\left(B+A\cosh{(\dual{X}/2)}\right)\right)
}{eq:gj18}
with the roles of $A,B$ interchanged. Indeed, there is an alternative action
\begin{equation}
  \label{eq:gj20}
  \dual{I}_1=\frac{1}{8\pi^2}\int\extd^2x\sqrt{-g}\left(\dual{X}R+\frac12\tanh{(\dual{X}/2)}(\nabla \dual{X})^2-\frac A4\sinh{\dual{X}}\right)\,,
\end{equation}
depending parametrically on $A$ with $B$ emerging as a constant of motion. The potentials are given by
\begin{equation}
  \label{eq:gj20aa}
  \dual{U}(\dual{X})=\frac12\tanh{(\dual{X}/2)}\,,\qquad \dual{V}(\dual{X})=\frac A4\sinh{\dual{X}}\,.
\end{equation}
\end{subequations}
Instead of a ``sine-Gordon'' potential, \eqref{eq:gj20} exhibits not only
a ``sinh-Gordon'' potential but also a kinetic term for the field $\dual{X}$. 
This formulation seems superior from a physical point of view for
three reasons: 1.~$\dual{X}\in[0,\infty)$ has
non-compact support. 2.~The constant of motion $B$ really plays the
physical role of the ADM mass (cf.~Section 5.1 in Ref.~\cite{Grumiller:2002nm}). 3.~The ground state solution $B=0$
is Minkowski space; actually, this is a consequence of the ``Minkowskian ground state
property'' property $\dual{V}e^{2\dual{Q}}\propto \dual{U}$ [cf.~(3.40) in Ref.~\cite{Grumiller:2002nm}].

The global structure of \eqref{eq:gj18} is very similar to the one of the Schwarzschild black hole: for $\dual{X}\to\infty$ geometry is flat and for $B+A\cosh{(\dual{X}/2)}=0$ a Killing horizon emerges. However, there is no curvature singularity at $\dual{X}=0$. For small values of the dilaton both actions asymptote to the Jackiw-Teitelboim model \cite{Jackiw:1984,Teitelboim:1984}.
%
Besides these features nothing noteworthy can be remarked about these geometries; there is no ``kink''-like structure, which was found in the $n=3\to n=2$ transition. 

\subsection{Intrinsically 2D dual Killing vector}

The 2D line element extracted from \eqref{eq:vkv17}
\begin{subequations}
\begin{equation}
  \label{eq:newlabel3}
  \extd s^2=2\extd u\extd X + \left(\frac{1}{4}X^4-\frac{Y}{2}X^2-2M\right)\extd u^2
\end{equation}
follows from the 2D dilaton gravity action
\begin{equation}
  \label{eq:conclusion1}
   I_2 = \frac{1}{8\pi^2}\int \extd^2x\sqrt{-g}\left(X R + YX-X^3\right)\,.
\end{equation}
[Here \eqref{eq:newlabel3} is presented with $A$ and $B$ replaced by
$-Y/2$ and $-2M$, respectively.]
Remarkably, this is almost identical to \eqref{eq:d1} in the following
sense: upon integrating out the vector field $a_\mu$ 
the {\em scalar field} $Y$ in \eqref{eq:d1} evidently is constant and
may be chosen to coincide with the {\em parameter} $Y$ in \eqref{eq:conclusion1}. The potentials are given by
\begin{equation}
  \label{eq:label17}
  U(X)=0\,,\qquad V(X)=YX-X^3\,.
\end{equation}
\end{subequations}

The global structure of the geometries \eqref{eq:newlabel3} has been analyzed already \cite{Grumiller:2003ad}; depending on the signs and magnitudes of $Y$ and $M$ there may be up to two Killing horizons in the region of positive $X$. Moreover, this model exhibits constant dilaton vacua, i.e., solutions where $X=\rm constant$, for $X=0$ and for $X=\pm\sqrt{Y}$. For positive $Y$ the solution attached to $X=0$ is $dS_2$ with $R=-Y$, while the ones attached to $X=\pm\sqrt{Y}$ are $AdS_2$ with $R=2Y$. The latter coincide with the solutions discussed in Section \ref{se:CDV} (because there $X$ has been rescaled to 1, which means $Y=1$ and thus reproduces \eqref{eq:const4} with $\la=1$). Thus, also the constant dilaton vacua are covered correctly by the action \eqref{eq:conclusion1}.
From a 2D point of view there has to be a kink-like solution interpolating between these constant dilaton vacua, by full analogy to \cite{Guralnik:2003we}. However, the vacuum $X=0$, although regular in 2D, is singular in three dimensions as the line element \eqref{eq:vkv17}
  degenerates for $X=0$. This explains the absence of kink-like solutions for the present case. 

For sake of completeness we mention that by using the potentials
\begin{subequations}
\begin{equation}
  \label{eq:label83}
  \dual{U}(\dual{X})=\frac{2}{\dual{X}}\,,\qquad \dual{V}(\dual{X})=-\frac{1}{2\dual{X}}-4M\dual{X}^3
\end{equation}
the alternative action
\begin{equation}
  \label{eq:dualnewaction}
   \dual{I}_2 = \frac{1}{8\pi^2}\int \extd^2x\sqrt{-g}\left(\dual{X} R + \frac{2}{\dual{X}}(\nabla{\dual{X}})^2+\frac{1}{2\dual{X}}+4M\dual{X}^3\right)\,
\end{equation}
again exchanges the respective roles of constant of motion and parameter of the action, leading to the line element
\begin{equation}
  \label{eq:le20}
   \extd s^2=\frac{1}{\dual{X}^2}\left(2\extd u\extd \dual{X} + \left(\frac{1}{4\dual{X}^2}-2M\dual{X}^2-\frac{Y}{2}\right)\extd u^2\right)\,.
\end{equation}
\end{subequations} 
The transformation $\dual{X}=-1/X$ brings \eqref{eq:le20} back into the form \eqref{eq:newlabel3}.

\section{Conclusions}

A Kaluza-Klein reduction of the $n$-dimensional conformal tensor (the Weyl tensor in $n\geq 4$) led to interesting equations \eqref{eq:j31} for conformal flatness of the $n$-dimensional space, which may be interpreted as equations of motion of some $n-1$ dimensional Einstein-Maxwell like theory. For $n=4$ drastic simplifications occurred \eqref{eq:j36}. The second of these equations, \eqref{eq:j36b}, exhibited the existence of a 3-dimensional Killing vector $f^\mu$, the dual field strength [cf.~\eqref{eq:j34}]. Depending on the properties of $f^\mu$ the existence of further Killing vectors in three dimensions could be shown, all of which can be lifted to four dimensions:
\blist
\item For $f^\mu=0$ the 3-dimensional space is maximally symmetric and thus has six Killing vectors.
\item For generic $f^\mu\neq 0$ and $d_\mu f_\nu$ not vanishing identically the existence of a dual Killing vector $F^\mu$ could be shown \eqref{eq:2ndkill2}.
\item For space-like $f^\mu\neq 0$ and $d_\mu f_\nu=0$ we could show that a further reduction to 2D is possible and that the 2-dimensional space is maximally symmetric. Therefore, in addition to $f^\mu$ there are three Killing vectors.
\item For time-like $f^\mu\neq 0$ and $d_\mu f_\nu=0$ one can repeat the analysis of the previous case and also ends up with three Killing vectors in addition to $f^\mu$.
\item The case of light-like $f^\mu\neq 0$ and $d_\mu f_\nu=0$ is not considered here.
\elist
So even in the most generic case there are in total three Killing vectors in the 4-dimensional theory: the first one is an input in the Kaluza-Klein {\em Ansatz} \eqref{eq:j1}, the second one, $f^\mu$, emerges from the equations of motion and the third one, $F^\mu$, is dual to the second one. This resembles somewhat the Schwarzschild situation while proving the Birkhoff theorem: with only the Killing vectors of spherical symmetry as input the vacuum Einstein equation implies another Killing vector related to staticity.

We proceeded to find all solutions based upon the {\em Ansatz} \eqref{eq:dd} and could reduce the system to 2D. All solutions fell into either of three classes depending on the properties of $f^\mu$: if it was covariantly constant we obtained maximally symmetric 2D sub-spaces \eqref{eq:solcons}; for an intrinsically 2D Killing vector we ended up with \eqref{sol1}; finally, if $f^\mu$ had no 2D components at all we were led to \eqref{sol2}. The dual Killing vector $F^\mu$ existed for the latter two cases and always was orthogonal to $f_\mu$, $F^\mu f_\mu=0$. All our 4-dimensional line elements were conformally flat by construction, but we have not presented a coordinate transformation which makes this manifest, $\extd s^2_{(4)}=e^{2\si}\eta_{MN}\extd x^M\extd x^N$. This purely technical excercise is not at all trivial. 

In the last part of our paper we constructed various 2D dilaton gravity actions, \eqref{eq:gj15}, \eqref{eq:gj20}, \eqref{eq:conclusion1} and \eqref{eq:dualnewaction}, which lead to equations of motion, the solutions of which reproduced the three classes mentioned above. This was remarkable insofar as there does not seem to be a straightforward way to obtain an action whose equations of motion are given by $C^{KLMN}=0$. The closest thing to such an action in the generic case is the well-known quadratic one,
\begin{subequations}
\begin{equation}
  \label{eq:wsq1}
  \int \extd^n x\sqrt{|g|} C_{KLMN}C^{KLMN}\,,
\end{equation}
which can be reduced to an $n-1$ dimensional action after inserting
\eqref{eq:j26} and integrating out the $n^{\rm th}$ coordinate. The ensuing equations of motion are rather complicated,
however, so we restrict ourselves to the simpler case $n=4$. The
reduced action (dropping the overall constant) 
\begin{equation}
  \label{eq:wsq2}
  \int \extd^3 x\sqrt{g} \left(c_{\mu\nu}c^{\mu\nu}-\frac14 K_{\mu\nu}K^{\mu\nu}\right)
\end{equation}
depends on the traceless symmetric tensors $c_{\mu\nu}$ introduced previously in \eqref{eq:j26b} and
\begin{equation}
  \label{eq:wsq4}
  K_{\mu\nu}=d_\mu f_\nu+d_\nu f_\mu\,.
\end{equation}
\end{subequations}
A special class of solutions is determined by the equations $c_{\mu\nu}=K_{\mu\nu}=0$, which are
identical to the conditions \eqref{eq:j36}. More general solutions are possible, but we have not attempted to construct them. Thus, this quadratic action encompasses all our solutions but also provides additional ones. 


Finally, we would like to comment on the relations between various 2D actions presented in this paper. The feature that two different actions describe the same set of classical solutions, but with the role of constant of motion and parameter in the action exchanged, has been a recurring theme. This property readily generalizes to a large class of dilaton gravity models and appears to be a fully fledged duality. It deserves further study. 

\section*{Acknowledgments}

DG would like to thank Alfredo Iorio for enjoyable discussions on gravitational Chern-Simons terms and their Kaluza-Klein reduction.

This work is supported in part by funds provided by the U.S. Department of Energy (D.O.E.) under the cooperative research agreement DEFG02-05ER41360.
DG has been supported by project GR-3157/1-1 of the German Research Foundation (DFG) and by the Marie Curie Fellowship MC-OIF 021421 of the European Commission under the Sixth EU Framework Programme for Research and Technological Development (FP6). 

\appendix

%

\section{Curvature in two and three dimensions}\label{app:2D}\label{app:B}

Here are details for 2D geometries of the form \eqref{eq:gj1}. The metric components read
\eq{
g_{uu}=e^QK\,,\qquad g_{uX}=e^Q\,,\qquad g_{XX}=0\,.
}{eq:2D1}
The Killing vector components are
\eq{
f^u=1\,,\qquad f^X=0\,,\qquad f_u=e^QK\,,\qquad f_X=e^Q\,.
}{eq:2D3}
The 2D Ricci scalar is given by
\eq{
R= V^\prime - 2UV - U^\prime e^{-Q} K  
}{eq:2D4}
where $U,V$ are defined in \eqref{eq:eqs}.
Derivatives of $X$,
\eq{
\square X = -V\,,\qquad (\nabla X)^2 = - e^{-Q} K
}{eq:2D5}
and
\[
\nabla_u\nabla_u\ln{X} = K \nabla_u\nabla_X\ln{X} = \frac{K}{2X}\left(KU-K^\prime\right)\,,\quad  \nabla_X\nabla_X\ln{X} = \frac{1}{X^2}\left(XU-1\right) 
\]
are needed to determine the 3-dimensional geometric quantities \eqref{eq:DD1} since $\phi$ is some function of $X$. In our problem
\begin{equation}
  \label{eq:neweq1}
  \phi=X^\al\,,\qquad \al=\frac12 \;{\rm or}\; 1\,.
\end{equation}
Thus, one obtains the 3-dimensional Ricci scalar
\begin{subequations}
\eq{
r=V^\prime+2V\left(\frac{\al}{X}-U\right)+e^{-Q}K\left(\frac{2\al(\al-1)}{X^2}-U^\prime\right)\,,
}{eq:2D9}
and the traceless portion of the 3-dimensional Ricci tensor ($\Rtr_{\mu\nu}:=r_{\mu\nu}-\frac13 g_{\mu\nu}r$)
\begin{align}
\Rtr_{uu}&= K\Rtr_{uX}\,,\\
\Rtr_{uX}&=\frac16 V^\prime e^Q-\frac16 Ve^Q\left(\frac{\al}{X}+2 U\right)+\frac16 K\left(\frac{4\al(1-\al)}{X^2}-U^\prime-\frac{3\al}{X}U\right)\,,\\
\Rtr_{XX}&=\frac{\al}{X^2}\left(1-\al-XU\right)\,,\\
\Rtr_{u\theta}&=\Rtr_{X\theta}=0\,,\\
\Rtr_{\theta\theta}&=\frac{X^{2\al}}{3} V^\prime-\frac{X^{2\al-1}}{3} V\left(\al+2XU\right)+\frac{X^{2\al-2}}{3} e^{-Q}K
 \left(\al(1-\al)-X^2U^\prime\right)\,.
\end{align}
\end{subequations}

\section{Tracefree combination of field strengths}\label{app:C}

For the Killing vector \eqref{eq:2D3} with $f^\theta=0$ the tracefree combination of field strengths reads ($\Ftr_{\mu\nu}:=f_\mu f_\nu-\frac13 g_{\mu\nu}f^2$)
\begin{subequations}
\begin{align}
\Ftr_{uu}&= K\Ftr_{uX}=\frac23 K^2 e^{2Q}\,,&\Ftr_{XX}&= e^{2Q} \,,\\
\Ftr_{u\theta}&=\Ftr_{X\theta}=0\,,&\Ftr_{\theta\theta}&=\frac{1}{3}X^{2\al} e^Q K \,,
\end{align}
\end{subequations}
where $\al$ is the exponent defined in \eqref{eq:neweq1}.
For the Killing vector $f^\theta=1$, $f^\al=0$ one obtains
\begin{subequations}
\begin{align}
\Ftr_{uu}&= K\Ftr_{uX}= \frac{1}{3} K e^Q X^{2\al} \,,&\Ftr_{XX}&= 0\,, \\
\Ftr_{u\theta}&=\Ftr_{X\theta}=0\,,&
\Ftr_{\theta\theta}&=\frac{2}{3}X^{4\al} \,.
\end{align}
\end{subequations}


\providecommand{\href}[2]{#2}\begingroup\raggedright\endgroup

\end{document}